\documentclass[aps,pra,reprint,superscriptaddress,floatfix]{revtex4-2}

\usepackage{amsmath,amssymb,graphicx}
\usepackage{bm}

\begin{document}

\title{Noise-driven pseudovorticity multipoles in self-focusing beams with quintic saturation}

\date{\today}
\author{Chengbo Zhang}
\author{Xiaohui Gao}
  \email{gaoxh@utexas.edu}
\affiliation{Department of Physics, Shaoxing University, Shaoxing, Zhejiang 312000, China}

\begin{abstract}
We investigate pseudovorticity generation in Gaussian beams undergoing self-focusing under amplitude and phase noise, using the cubic-quintic nonlinear Schr\"odinger equation. Pseudovorticity, defined as the curl of the optical momentum flux, characterizes local rotational flow in the absence of phase singularities. Our numerical simulations show that thermal amplitude and phase noise induce a multipolar pseudovorticity pattern. Unlike the pure cubic case, where noise asymmetries are radiated away during collapse, the quintic saturation arrests collapse and traps the noise in the resulting soliton. Hence, pseudovorticity multipoles persist, oscillating at the focusing-refocusing period. These results suggest a potential pathway for controlling local optical torque through noise engineering.
\end{abstract}

\maketitle

\section{Introduction}
Kerr-induced self-focusing occurs when a positive nonlinear refractive index, $n = n_0 + n_2 I$, creates an effective focusing lens~\cite{Marburger1975PQE,Couairon2007PR}. This process is modeled by the nonlinear Schrödinger equation (NLSE), a framework that appears across diverse fields, including nonlinear optics, hydrodynamics, quantum fluids, and plasma physics~\cite{fibich_nonlinear_2015}. Well-known features emerging from this model include the universal blowout Townes profile~\cite{Moll2003PRL} and the critical power for catastrophic self-focusing~\cite{Fibich2000OL}. Variations in initial beam shapes, such as Gaussian versus super-Gaussian profiles~\cite{Grow2006OE}, or in the underlying saturation mechanisms~\cite{MihalachePRE2003,AbdullaevPRE05, Pasquazi2014}, give rise to a diverse range of dynamical behaviors.

As real-world beams are inevitably subject to thermal fluctuations, quantum noise, and fabrication imperfections, the effect of the noise is important. Depending on the power, the influence of such noise on NLSE dynamics broadly falls into two regimes~\cite{Zhang25}. The first is optical self-cleaning~\cite{Moll2003PRL,Prade2006,Liu2007}, where asymmetries are radiated outward from the high-intensity core into a low-intensity reservoir, leaving the core symmetric and erasing the memory of the initial perturbations~\cite{Sagiv2020PD}. The second is modulation-instability-induced filamentation~\cite{Fibich2001,Garnier2006}, in which noise seeds localized hot spots, and the statistical properties of the input, such as the coherence length, determine the output characteristics~\cite{ChoudharyPRR2024}.

Pseudovorticity, also referred to as current vorticity~\cite{Berry2009JA}, has recently attracted interest~\cite{Falsi2026} as it gives rise to local rotational structures even in the absence of phase singularities, and is relevant to applications such as optically induced torque on small particles~\cite{Roichman2008PRL}. Experiments in photorefractive media show that 2+1D optical solitons naturally host pseudovorticity~\cite{Falsi2026}. For an ideal Gaussian beam without noise or anisotropy term, the field retains cylindrical symmetry throughout propagation, and the pseudovorticity remains zero. As the pseudovorticity is proportional to spatial gradients of both intensity and phase, one may expect that it is exquisitely sensitive to noise.

In this work, we study the propagation dynamics of noise-driven pseudovorticity in a Gaussian beam with noise using the cubic-quintic nonlinear Schr\"odinger equation, which introduces a saturating nonlinearity. We find that the beam does not evolve into a featureless Townes profile or chaotic hot spots; instead, specific components, particularly quadrupolar modes, are captured as bound internal states~\cite{MihalachePRE2003} and persist within the fundamental soliton. This trapping gives rise to sustained oscillatory pseudovorticity multipoles, offering a potential pathway to control local optical torque through input noise engineering.

\section{Theoretical Model and Methods}
Our analysis is based on the (2+1)-dimensional cubic-quintic nonlinear Schr\"odinger equation (NLSE), written in normalized form as~\cite{Sagiv2020PD}
\begin{equation}
i\frac{\partial\psi}{\partial z} + \nabla_\perp^2 \psi + |\psi|^2\psi - \epsilon |\psi|^4\psi = 0,
\end{equation}
where $\psi$ denotes the slowly varying envelope normalized to the initial peak field amplitude, $z$ is the propagation coordinate scaled to the diffraction length of the input beam, and $(x,y)$ are the transverse coordinates normalized to the initial $1/e$ field radius. The transverse Laplacian $\nabla_\perp^2 = \partial_x^2 + \partial_y^2$ accounts for diffraction. The cubic term describes the conventional Kerr self-focusing, while the negative quintic term models the saturating higher-order nonlinearity. This phenomenological model provides an effective description of beam dynamics in various saturable media, including doped glasses~\cite{Senthilnathan08PRA} and photorefractive crystals~\cite{jeng_partially_2004,boyd_self-focusing_2009-1}. Here we take $\epsilon = 10^{-2}$, a value that yields a soliton of moderate transverse extent, facilitating good spatial resolution on our numerical grid.  As the incident condition, we launch a Gaussian beam $\psi_0 = 6\exp(-r^2)$ with $r^2 = x^2 + y^2$, carrying an input power of approximately $3.3P_{\mathrm{cr}}$, where $P_{\mathrm{cr}}$ denotes the critical power for catastrophic self-focusing.

Consider a complex scalar field $\psi = \sqrt{I}e^{i\phi}$, where $I = |\psi|^2$ denotes the optical intensity and $\phi$ is the phase. In the paraxial scalar approximation, the transverse momentum density (i.e., the optical current) is defined as
\begin{equation}
    \mathbf{j} = I \nabla_{\!\perp} \phi = \operatorname{Im}(\psi^* \nabla_{\!\perp} \psi).
    \label{eq:current}
\end{equation}
The pseudovorticity is then defined as the curl of this momentum density:
\begin{equation}
    \omega = \nabla_{\!\perp} \times \mathbf{j} = \nabla_{\!\perp} I \times \nabla_{\!\perp} \phi.
    \label{eq:ps_def}
\end{equation}
In two-dimensional Cartesian coordinates $(x,y)$, this explicitly reads
\begin{equation}
    \omega = \frac{\partial I}{\partial x}\frac{\partial \phi}{\partial y} - \frac{\partial I}{\partial y}\frac{\partial \phi}{\partial x},
    \label{eq:ps_cart}
\end{equation}
while in polar coordinates $(r,\theta)$, the equivalent expression is
\begin{equation}
    \omega = \frac{1}{r}\left( \frac{\partial I}{\partial r}\frac{\partial \phi}{\partial \theta} - \frac{\partial I}{\partial \theta}\frac{\partial \phi}{\partial r} \right).
    \label{eq:ps_polar}
\end{equation}
Equivalently, pseudovorticity can be expressed directly in terms of the complex field $\psi$ as
\begin{equation}
    \omega = \nabla_{\!\perp} \times \operatorname{Im}(\psi^* \nabla_{\!\perp} \psi) = \operatorname{Im}\left[\nabla_{\!\perp} \psi^* \times \nabla_{\!\perp} \psi\right],
    \label{eq:ps_psi}
\end{equation}
which highlights its connection to the phase-amplitude coupling of the wavefield.

This quantity measures the local inhomogeneity of the optical current and is nonzero whenever the intensity and phase gradients are non-collinear. Within the paraxial scalar approximation, the local optical torque density exerted on a small absorbing particle is proportional to the pseudovorticity integrated over the particle's cross-section. Thus, pseudovorticity provides a direct and powerful diagnostic of rotation-inducing field structures, even in the absence of phase singularities or net orbital angular momentum.

\begin{figure}[htbp]
\centering
\includegraphics[width=0.45\textwidth]{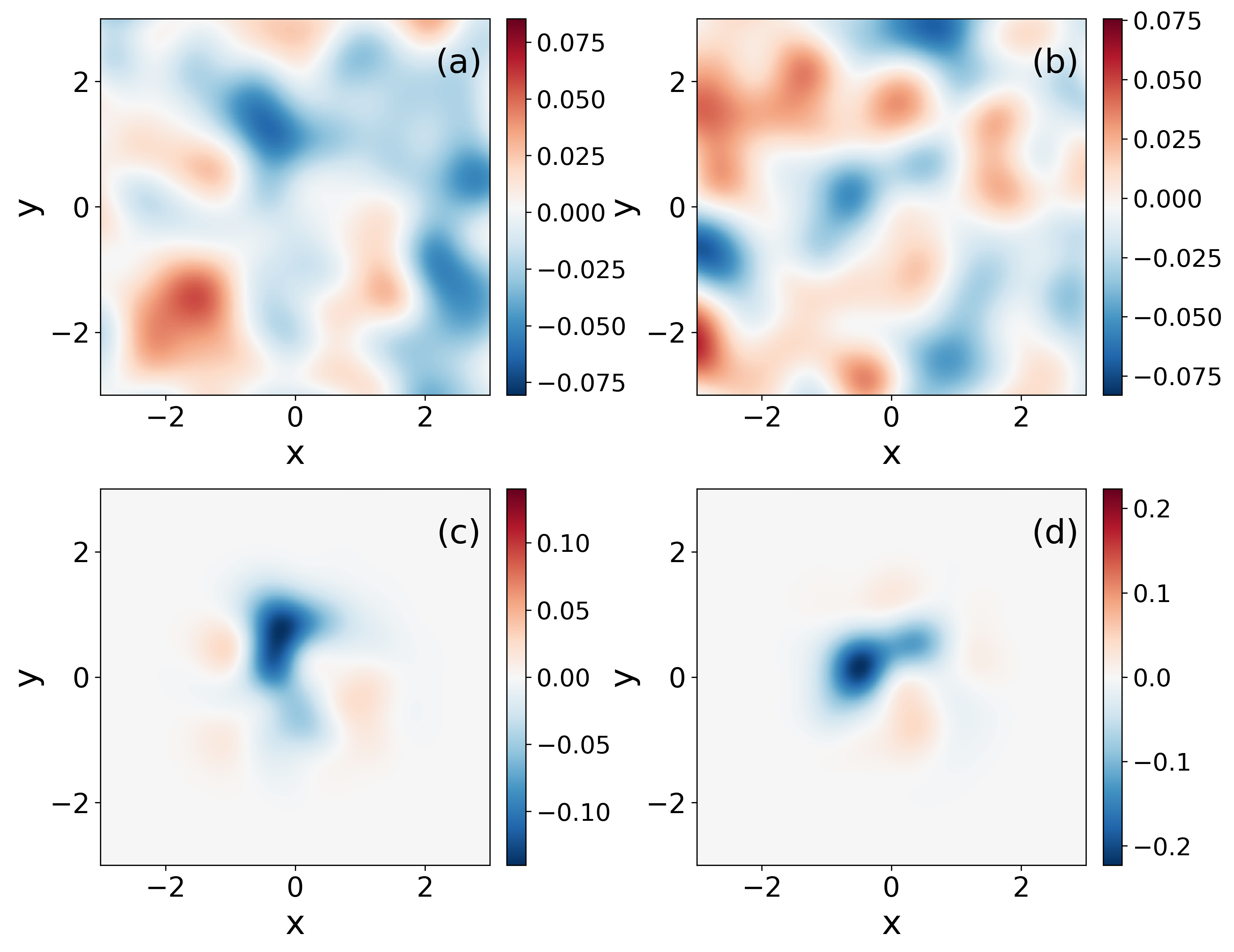}
\caption{Initial simulation conditions with thermal noise correlation length \( \sigma = 20\Delta x \) and RMS amplitude \(0.02\). (a), (b) Thermal noise field \( \eta(x, y) \) for random seeds 0 and 1, respectively. (c), (d) Corresponding initial noisy component \( \psi_0\eta_A \) for seeds 0 and 1, respectively.}
\label{fig1}
\end{figure}%
As real noise fields are not completely random and have coherence length, amplitude and phase noise are modeled using a thermal noise approach. Gaussian white noise is generated and then smoothed with a Gaussian filter to introduce spatial correlation:
\begin{equation}
\eta = G_\sigma * \mathcal{N}(0,1),
\label{eqeta}
\end{equation}
where $G_\sigma$ is the Gaussian kernel with standard deviation $\sigma$, and $\mathcal{N}(0,1)$ represents normally distributed random numbers, and $*$ denotes convolution. The amplitude noise is normalized to a specified root-mean-square (RMS) level, and the phase noise is normalized to a specified RMS value in radian. The noisy field is:
\begin{equation}
\psi_{\mathrm{noisy}} = \psi_0 (1 + \eta_A) e^{i\eta_P},
\end{equation}
where $\eta_A$ and $\eta_P$ are the noise generated using Eq.~\ref{eqeta} for amplitude and phase noise, respectively. The field are rescaled to keep the power the same. For amplitude noise, we use $\mathrm{RMS} = 0.02$ and $\sigma = 20\Delta x = 0.38$. For phase noise, we use $\mathrm{RMS} = 0.1$ rad. These noise levels are routinely encountered in laboratory settings. Here representative examples of the amplitude noise fields $\eta(x,y)$ and the resulting initial perturbations $\psi_0\eta_A$ for two different random seeds are displayed in Fig.~\ref{fig1}. With the initial conditions, numerical integration is performed using the split-step Fourier method with grid size $1024\times1024$, spatial window half-width $X_{\max}=20$, and step size $\Delta z = 0.001$.

\section{Numerical Results}
\begin{figure}[htbp]
\centering
\includegraphics[width=0.40\textwidth]{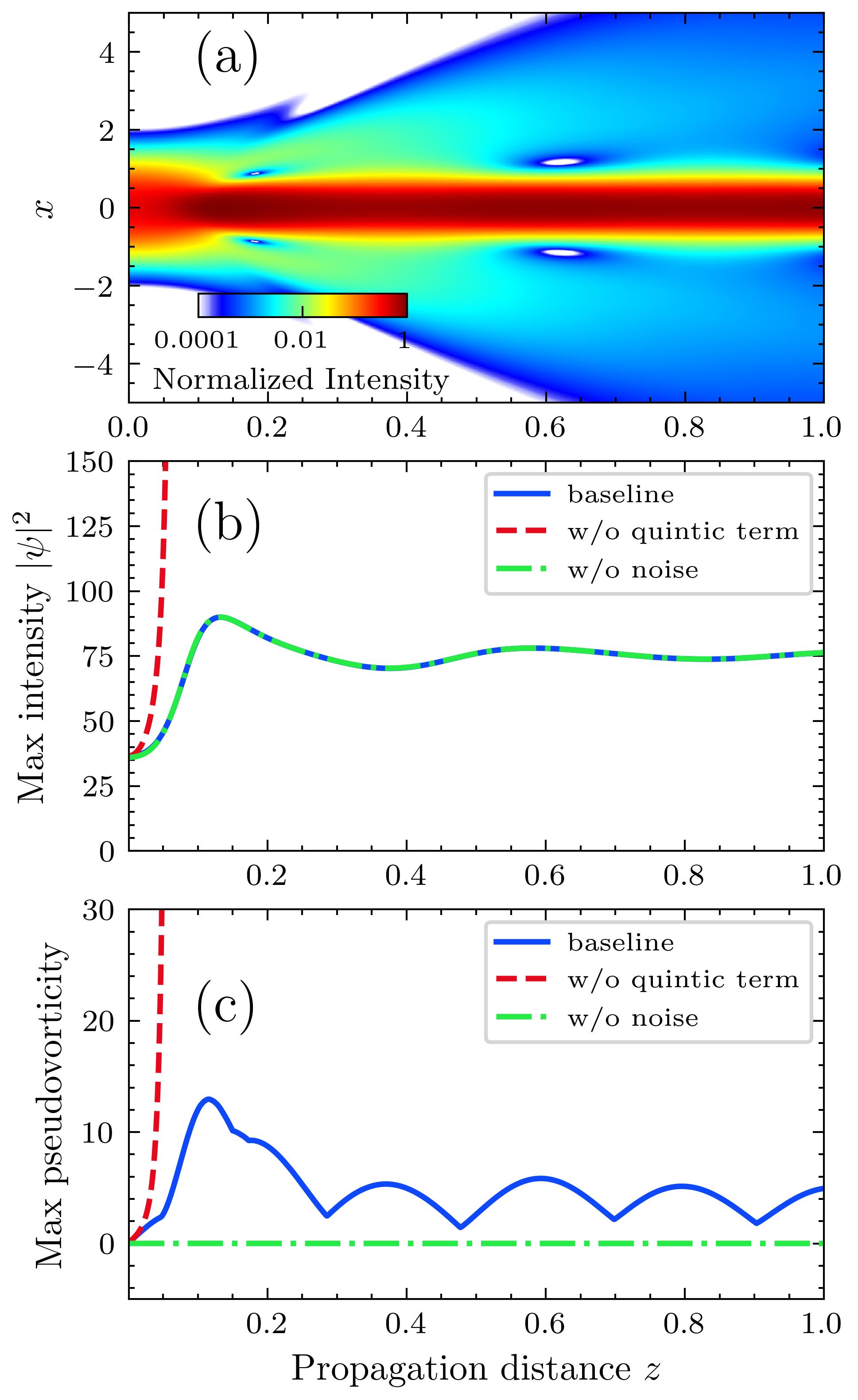}
\caption{Amplitude noise results ($\sigma=20\Delta x$, RMS $0.02$, seed $0$). (a) Transverse intensity profile $|\psi(x,y=0)|^2$ versus $z$. (b) Maximum intensity and (c) maximum pseudovorticity as functions of $z$. Blue solid: cubic-quintic with noise; red dashed: pure cubic with noise; green dash-dotted: cubic-quintic without noise.}
\label{fig2}
\end{figure}

We begin by examining amplitude thermal noise as a representative perturbation. Figure~\ref{fig2} presents results for RMS $0.02$, correlation length $\sigma = 20\Delta x$, and seed $0$. In Fig.~\ref{fig2}(a), the transverse intensity profile $|\psi(x,y=0)|^2$ versus $z$ for the cubic-quintic model with noise shows that collapse is arrested at $z \approx 0.13$; the beam subsequently defocuses and refocuses, settling into a persistent soliton state, while the low-intensity background expands monotonically with propagation. 
Figures~\ref{fig2}(b) and~\ref{fig2}(c) plot the transverse maxima of intensity ($\max|\psi|^2$) and pseudovorticity ($\max|\omega|$), respectively, as functions of $z$, comparing three cases: the cubic-quintic model with noise (blue solid), the pure cubic model with the same noise (red dashed), and the cubic-quintic model without noise (green dash-dotted). The green curve nearly overlaps with the blue one for the intensity maximum, indicating that noise has minimal effect on the overall intensity dynamics. As expected, the noise-free case yields zero pseudovorticity throughout, confirming numerical accuracy.
For the noisy cubic-quintic case, the pseudovorticity maximum exhibits a sharp initial spike of about $13$ during the first collapse; it then decreases and enters a sustained oscillatory regime with a period of $\Delta z \approx 0.2$, which is half the focusing–refocusing period of the intensity breathing. The frequency doubling arises from taking the absolute value. The first oscillation cycle has a substantially higher amplitude and contains minor substructure, while the subsequent cycles are cleaner and more regular. This reflects that the major beam reshaping occurs during the initial collapse. In the pure cubic model, both the intensity and pseudovorticity maxima display steep, unbounded growth, consistent with critical collapse. Theoretically, the pseudovorticity should eventually decay through self-cleaning, as the noise is expelled from the shrinking core; however, our simulation cannot resolve this decay due to insufficient spatial resolution during the late stages of collapse.

\begin{figure}[htbp]
\centering
\includegraphics[width=0.45\textwidth]{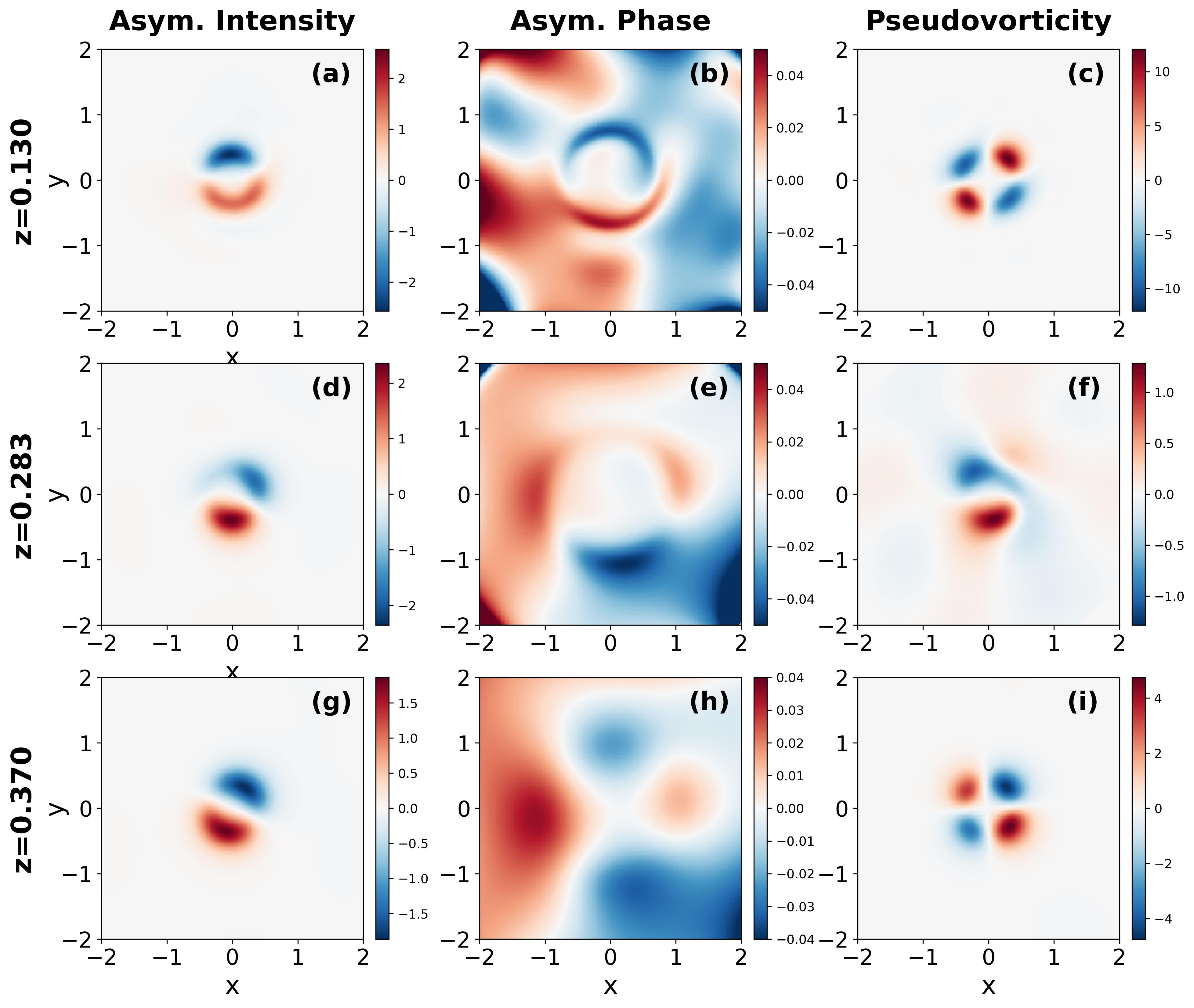}
\caption{Snapshots at focus ($z=0.13$), mid-focus ($z=0.283$), and second focus ($z=0.37$) (rows). Columns: asymmetric intensity (left), asymmetric phase (middle), pseudovorticity (right).}
\label{fig3}
\end{figure}%
Figure~\ref{fig3} shows spatial snapshots at three characteristic propagation distances: the first focus ($z=0.13$), a middle position ($z=0.283$), and the second focus ($z=0.37$). The intensity and phase are decomposed into their symmetric (angularly averaged) and asymmetric parts via the Fourier series
\begin{align}
I(r,\theta) &= I_0(r) + \sum_{m=1}^{\infty} \left[ I_m^{(c)}(r)\cos(m\theta) + I_m^{(s)}(r)\sin(m\theta) \right], \\
\phi(r,\theta) &= \phi_0(r) + \sum_{m=1}^{\infty} \left[ \phi_m^{(c)}(r)\cos(m\theta) + \phi_m^{(s)}(r)\sin(m\theta) \right],
\end{align}
where $m$ is a non-negative integer denoting the angular harmonic order: $m=0$ corresponds to the cylindrically symmetric (mean) component, $m=1$ to the dipole, $m=2$ to the quadrupole, $I_0(r)$ and $\phi_0(r)$ are the $m=0$ (cylindrically symmetric) components. Each row in Fig.~\ref{fig3} displays the asymmetric intensity residue $|\psi|^2 - |\psi|_{\mathrm{sym}}^2$ (left), the asymmetric phase $\phi-\phi_{\mathrm{sym}}$ (middle), and the corresponding pseudovorticity $\omega$ (right). The intensity residue remains appreciably nonzero with a dipole-like pattern, arising because the noise breaks the input symmetry, shifting the beam centroid and producing a positive–negative lobe structure. The phase remains continuous, confirming the absence of phase singularities. Since the intensity gradient vanishes at the beam center, the pseudovorticity there is small; it is instead localized at interfaces where $\nabla I$ and $\nabla \phi$ are non-collinear. As the beam breathes, the pseudovorticity pattern appears to alternate between quadrupolar and dipolar symmetries.

\begin{figure}[htbp]
\centering
\includegraphics[width=0.45\textwidth]{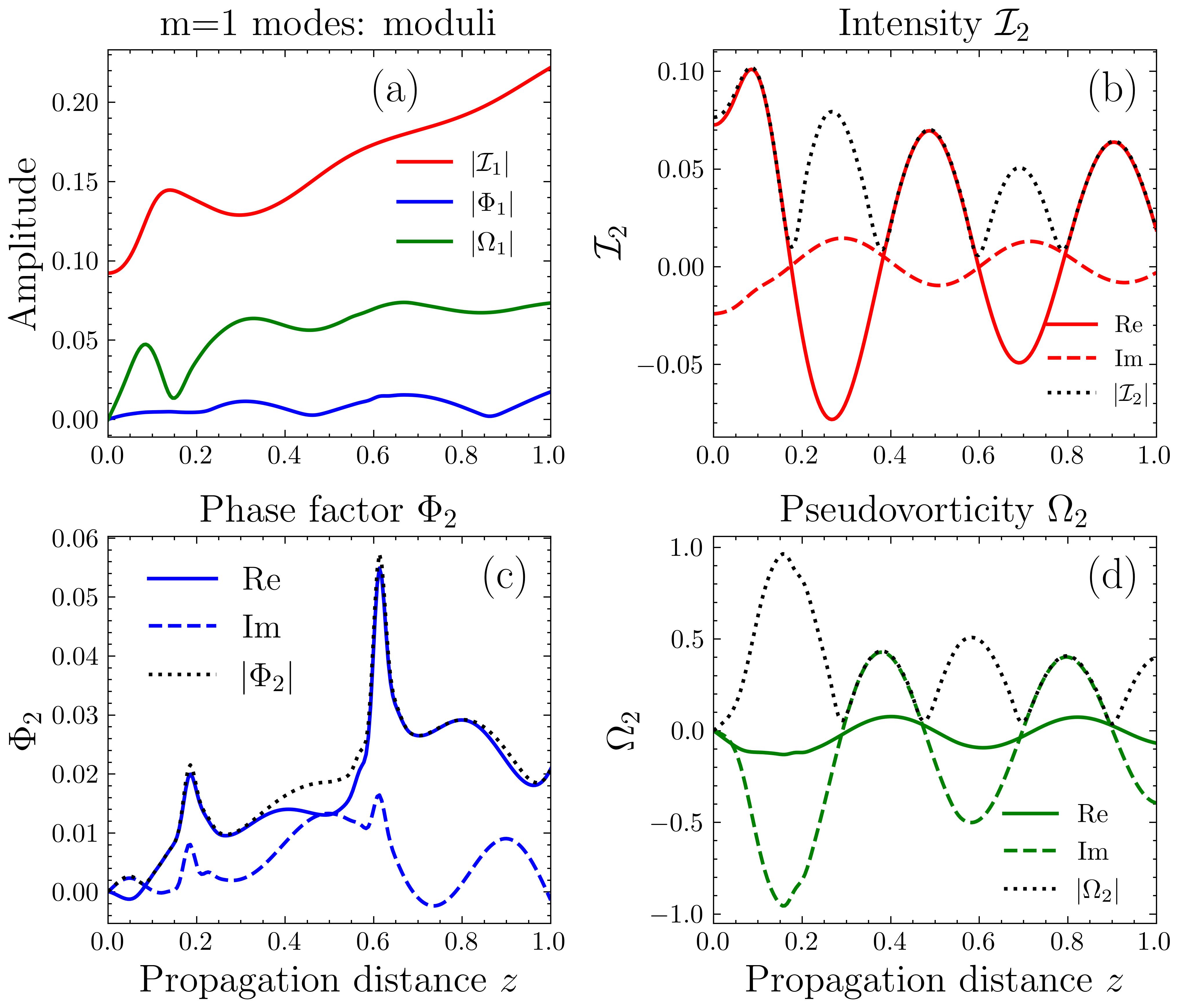}
\caption{(a) Moduli of dipole ($m=1$) components. (b)--(d) Real (solid), imaginary (dashed), and modulus (dotted) of quadrupole ($m=2$) components for intensity, phase factor, and pseudovorticity.}
\label{fig4}
\end{figure}%
To quantify the evolution of the global angular structure during propagation, we decompose the intensity $I=|\psi|^2$, the phase factor $e^{i\phi}$, and the pseudovorticity $\omega$ into angular harmonics and then integrate the modal coefficients over the transverse plane. The corresponding global $m$-th modal coefficients are defined as
\begin{align}
\mathcal{I}_m(z) &= \int_0^{R_{\max}}  \int_0^{2\pi} I(r,\theta,z) e^{-im\theta} d\theta  r \, dr, \\
\Phi_m(z) &= \int_0^{R_{\max}}  \int_0^{2\pi} e^{i\phi(r,\theta,z)} e^{-im\theta} d\theta  r \, dr, \\
\Omega_m(z) &= \int_0^{R_{\max}}  \int_0^{2\pi} \omega(r,\theta,z) e^{-im\theta} d\theta  r \, dr,
\end{align}
with $m=1$ (dipole) and $m=2$ (quadrupole). We use $R_{\max}=2$. For the phase, we use $e^{i\phi}$ to avoid complications from phase unwrapping. Figure~\ref{fig4}(a) shows the moduli of the dipole components $\mathcal{I}_1$, $\Phi_1$, and $\Omega_1$; the evolution of $|\mathcal{I}_1|$ is consistent with the shifting of the beam centroid (not shown here). 
Figures~\ref{fig4}(b)--(d) display the real, imaginary, and modulus parts of the quadrupole components $\mathcal{I}_2$, $\Phi_2$, and $\Omega_2$, respectively. The intensity quadrupole is predominantly real, while the pseudovorticity quadrupole is overwhelmingly imaginary; as a result, their oscillations are shifted by approximately $\pi/2$. Overall, the quadrupole mode exhibits a larger amplitude and stronger oscillations than the dipole. Near the troughs of the quadrupole oscillation, the dipole mode becomes dominant, giving rise to the transient dipolar structure observed in Fig.~\ref{fig3}(f).

\begin{figure}[htbp]
\centering
\includegraphics[width=0.45\textwidth]{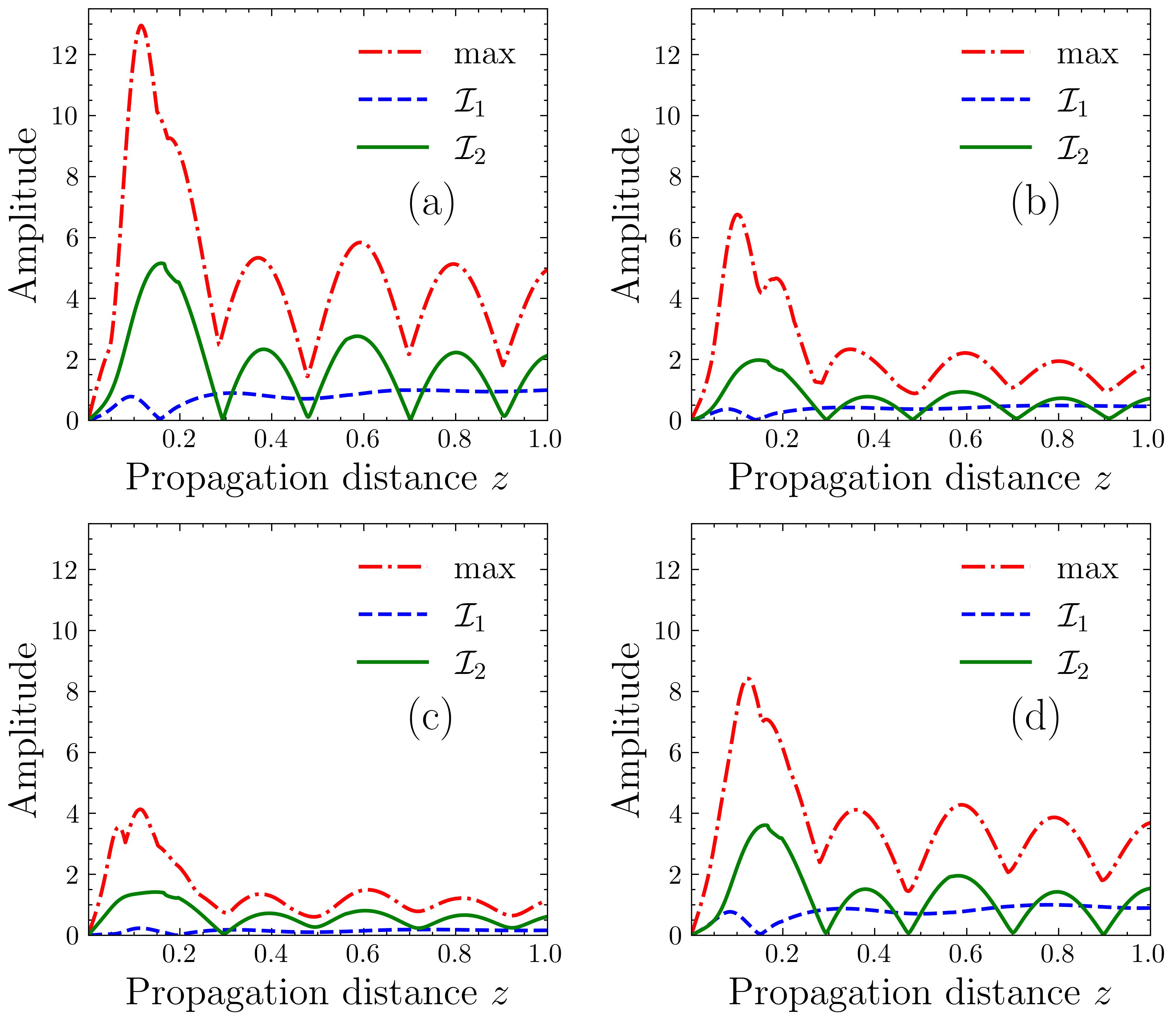}
\caption{Dipole, quadrupole, and peak pseudovorticity versus $z$ for seeds 1--4 (panels (a)--(d)) with amplitude noise $\sigma=20\Delta x$, RMS $0.02$.}
\label{fig5}
\end{figure}

To test the robustness against stochastic fluctuations, four independent noise realizations (seeds 1--4) with identical amplitude noise parameters were performed. Figure~\ref{fig5} presents the dipole coefficient $\mathcal{I}_1$, the quadrupole coefficient $\mathcal{I}_2$, and the maximum pseudovorticity versus $z$ for each seed. The maximum is a local quantity, while the coefficients are radially integrated modal amplitudes; their magnitudes should therefore not be compared directly. Despite seed-to-seed variations in the initial spike height (ranging from about 4 to 13 in the pseudovorticity maximum), all realizations exhibit common qualitative features: a strong initial spike near $z\approx 0.13$, followed by regular quasi-periodic oscillations with reduced amplitude, and a general dominance of the quadrupole over the dipole except at the oscillation troughs. These common features provide evidence that the observed multipolar oscillations are generic and not due to a specific noise seed.

\begin{figure}[htbp]
\centering
\includegraphics[width=0.45\textwidth]{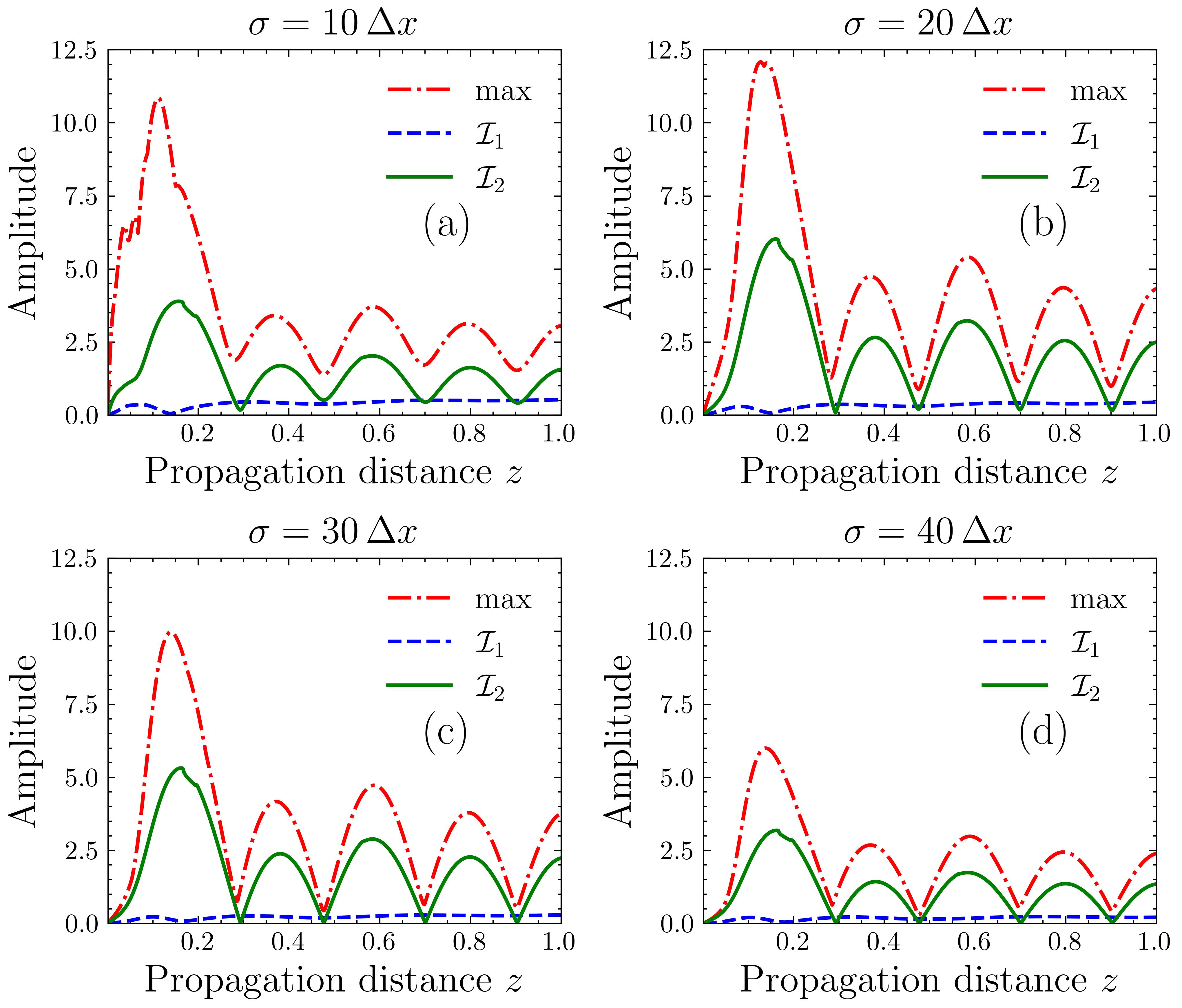}
\caption{Dipole, quadrupole, and peak pseudovorticity versus $z$ for amplitude noise with $\sigma = 10\Delta x, 20\Delta x, 30\Delta x, 40\Delta x$ (panels (a)--(d)).}
\label{fig6}
\end{figure}

The spatial correlation length of the amplitude noise was varied across $\sigma = 10\Delta x, 20\Delta x, 30\Delta x$, and $40\Delta x$. Figure~\ref{fig6} displays that the quadrupole and maximum pseudovorticity exhibit continuing oscillations with a period equal to half the soliton breathing cycle, and the quadrupole generally dominates the dipole. The case $\sigma = 20\Delta x$ produces the largest oscillation amplitudes, possibly because this correlation length matches the transverse extent of the initial beam best. Apart from this amplitude variation, the mode competition dynamics remain qualitatively similar across all correlation lengths, which shows the robustness of the trapping mechanism against changes in the amplitude noise correlation length.

\begin{figure}[htbp]
\centering
\includegraphics[width=0.45\textwidth]{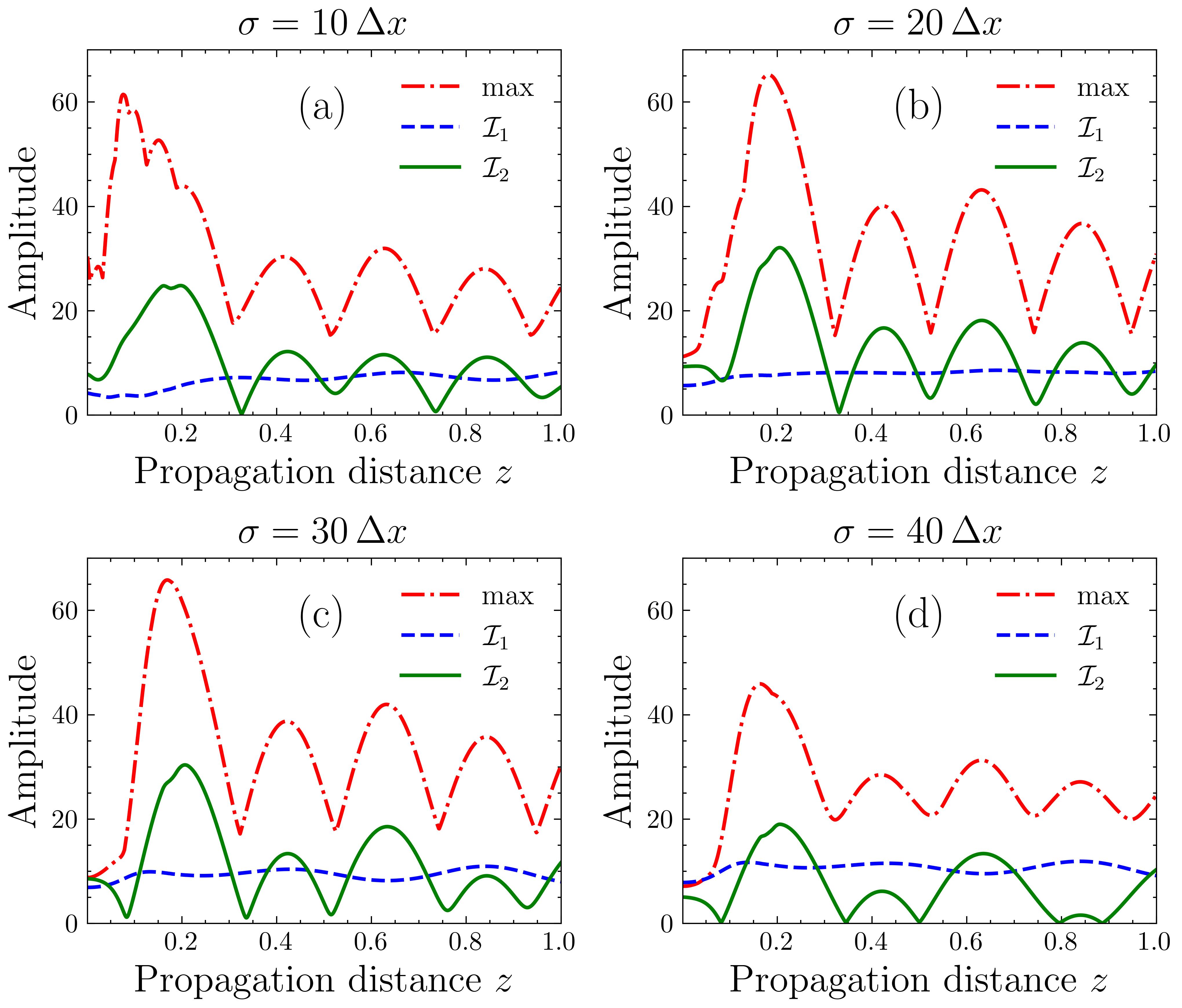}
\caption{Dipole, quadrupole, and peak pseudovorticity versus $z$ for phase noise with $\sigma = 10\Delta x, 20\Delta x, 30\Delta x, 40\Delta x$ (panels (a)--(d)).}
\label{fig7}
\end{figure}

Finally, phase noise with RMS $0.1$ rad was examined for the same set of correlation lengths. As illustrated in Fig.~\ref{fig7}, the oscillatory behavior remains evident, which verifies that the quintic saturation traps noise regardless of its type. The pseudovorticity amplitudes are roughly an order of magnitude larger than in the amplitude noise case, although a direct quantitative comparison is not meaningful because amplitude and phase noise have different physical units. 
For phase noise, the dipole and quadrupole coefficients become more comparable in magnitude, with the dipole coefficient being more stable while the quadrupole coefficient exhibits stronger oscillations. As a result, the dipole mode tends to dominate over a larger fraction of the propagation distance, particularly for longer correlation lengths ($40\Delta x$). This indicates that smoother phase perturbations preferentially couple energy into the dipole channel, in contrast to the amplitude noise case where the quadrupole consistently dominates. Such long-correlated phase structures are characteristic of common optical aberrations, such as coma, astigmatism, spherical aberration, and defocus, which typically have coherence lengths larger than the beam size.

\section{Conclusion}
In summary, we have numerically studied pseudovorticity generation in self-focusing Gaussian beams subject to amplitude and phase noise within the cubic-quintic NLSE. The quintic nonlinearity arrests collapse, allowing the beam to propagate as a breathing soliton. Noise-induced asymmetries are not radiated away but persist as internal modes, giving rise to sustained periodic oscillations of the pseudovorticity multipoles. Angular mode decomposition reveals that the quadrupole generally dominates the dipole, with transient reversals at oscillation troughs, a behavior reproducible across multiple noise seeds. These results establish quantitative links between input noise statistics and the multipolar rotational structure, suggesting that local optical torque can be controlled by tailoring noise properties in saturable media.

\end{document}